\documentclass[%
aip,
rsi,%
amsmath,amssymb,
reprint,%
]{revtex4-1}

\usepackage{graphicx}
\usepackage{dcolumn}
\usepackage{bm}
\usepackage{algorithm}%
\usepackage{algorithmicx}%
\usepackage{algpseudocode}%
\usepackage{float}
\usepackage[colorlinks=true, linkcolor=blue, citecolor=blue, urlcolor=blue]{hyperref}

\begin{document}
	
	\title{The thin line for optical neural networks towards broad practical relevance}
	
	\author{Anas Skalli}
	\affiliation{Universit\'{e} Marie et Louis Pasteur, CNRS UMR 6174, Institut FEMTO-ST, 15B Avenue Montboucons, Besançon, 25000, France}

	\author{Daniel Brunner}
	\email{daniel.brunner@femto-st.fr}	
	\affiliation{Universit\'{e} Marie et Louis Pasteur, CNRS UMR 6174, Institut FEMTO-ST, 15B Avenue Montboucons, Besançon, 25000, France}
	
	\date{\today}
	
	\begin{abstract}
		
Optical neural networks promise unmatched efficiency, bandwidth, and latency, critical benefits as demand for neural network hardware surges.  However, their practical value for general-purpose acceleration or specialized applications must be proven under application-realistic conditions.  We discuss recent insights and outline key research priorities.
		
	\end{abstract}
	
	\maketitle
	
\section{Introduction}

Once again \cite{Farhat1985optical} optical neural networks (ONNs) stand at the forefront of technological innovation, with now more than a decade into a renewed cycle of interest sparked by photonic reservoir computing \cite{larger2012photonic} and integrated ONNs \cite{shen2017deep}.
 What distinguishes today from before is a societal and economic demand for ANNs colliding with a lack of a clear and long-term sustainable ANN hardware vision.
 Despite their potential stemming from superior parallelism and signal transduction compared to artificial neural networks (ANNs) on electronic systems, real-world deployment remains hindered by technical hurdles and uncertainties, rendering success outside research labs nearly elusive.

The potential of AI systems that harness light is widely recognized and grounded in fundamental physical principles \cite{mcmahon2023physics}, and recent years have indeed witnessed astonishing advances.
 However, despite numerous years of support, the field has yet to unlock its full potential and continues to require long term development to reach real-world relevance.
 The scale of this challenge is frequently underestimated in research contexts.
 As John Maynard Keynes observed: \emph{It is not sufficient that the state of affairs which we seek to promote should be better than the state of affairs which preceded it; it must be sufficiently better to make up for the evils of the transition}.
 And, closer to our specific topic, the unexpectedly large challenge of moving optical computing from demonstrators to deployable systems as formulated by John Caulfield in ‘Perspectives on optical computing’ back in 2002.
 With these considerations in mind, this commentary examines ONNs in terms of potential future application categories.
 
The challenge outlined by Maynard Keynes of inducing broad-scale transformation and their combination with the intricacies of computing outlined by Caulfield stretch the path from proof-of-concept ONNs towards practical relevance into a thin line.
 It is however very likely that the indisputable necessity and potential benefits justify long- term and substantial commitment of resources.
 Such commitment over a long duration ultimately relies on trust. Trust from investors, private or public, into claims made by the scientific community.
 And trust of the scientific community that this support merits commitment to long-term research that does not shy away from addressing the real challenges.
 In this commentary, we examine general-purpose ANN acceleration and specialized ONNs designed for specific applications.
 These can either be fully implemented ONNs or optical sub-systems to partially accelerate ANN computations to achieve superior performance in some relevant metric.
 As these approaches introduce distinct and often orthogonal key performance indicators, they each necessitate tailored evaluations that account for the unique challenges of each.

\section{ONNs for ANN acceleration or physical computing}

General-purpose ANN computing enjoys widespread relevance, yet continuously improving mature solutions exist, even if they do under-perform to match ANN needs and growth rates.
 As a result, for real-world relevance optical ANN accelerators must reach an exceptional level of deployability.
 However, success would justify substantial research investments due to the evident transformative potential.
 In contrast, application specialized ONN target breakthroughs by entering regimes that are out-of-reach for other approaches, an ONN advantage in the context of quantum computing so to speak \cite{mcmahon2023physics}.
 They potentially face lower technical barriers due to the absence of direct competition, yet their success may hinge on identifying or even creating niche-use cases, presumably within smaller market segments. 
 
The main ANN computing performance bottleneck is inherently tied to its core structure: neurons are organized in successive layers to form deep networks. Before applying nonlinearity, each neuron computes weighted sums of its inputs.
 This multiply-accumulate (MAC) operation corresponds to matrix–vector multiplication (MVM), and for a comparable number of neurons N in all layers, the number of linear operations required to densely connect two layers scales quadratically, $O(N^2)$.
 Since nonlinear activations scale only linearly, $O(N)$, MVM operations dominate the computational load in ANNs.

Implementing these connections through physical, typically analogue, pathways is known as in-memory computing.
 Furthermore, nonvolatile in-memory computing can reduce data-movement overhead and hence mitigate the famous von Neumann bottleneck.
 Optical MVM is particularly promising, as it enables in principle large numbers of parallel MAC operations at high bandwidth, near-zero latency, extremely low energy per operation, potentially outperforming electronic interconnects.
 Consequently, both general-purpose optical ANN accelerators and specialized ONNs seek to exploit the advantages of optical MVM.

In-memory computing for ANN acceleration faithfully maps a classical ANN model onto the MVM circuit’s topology, and its hardware therefore must accurately implement the required MVMs.
 For ANN connection layers exceeding the hardware’s MVM dimension, the weight matrix is tiled into smaller operations (Fig. 1A).
 This enables MVM engines to be applied to ANNs varying both in size and depth, making optical accelerators highly impactful, as they can implement state-of-the-art ANN architectures trained in silico using standard machine-learning tools, including error backpropagation \cite{ahmed2025universal}.
 Promising platforms for such general-purpose photonic ANN acceleration include Mach–Zehnder meshes \cite{shen2017deep}, ring-resonator networks \cite{tait2017neuromorphic}, and photonic memristive crossbar arrays \cite{feldmann2021parallel}.
 In contrast, physical-computing–inspired ONN architectures compute by directly leveraging the hardware’s fundamental transformations, often incorporating optical nonlinearities.
 Representative examples include optical reservoir computing \cite{larger2012photonic}, spiking ONNs \cite{hurtado2012investigation}, and autonomous ONNs with minimal pre- or post-processing \cite{skalli2025model}.
 Their goal is to perform computation directly in the optical domain to maximally leverage scaling, speed, and efficiency advantages at the system level rather than only at the level of individual MVMs.
 In this context, ONNs do not aim for mathematical isomorphism with an ANN model, but ANNs provide inspiration for ONN topology and data-driven optimization \cite{momeni2025training}.

\begin{figure*}[t]
	\centering
	\includegraphics[width=1\linewidth]{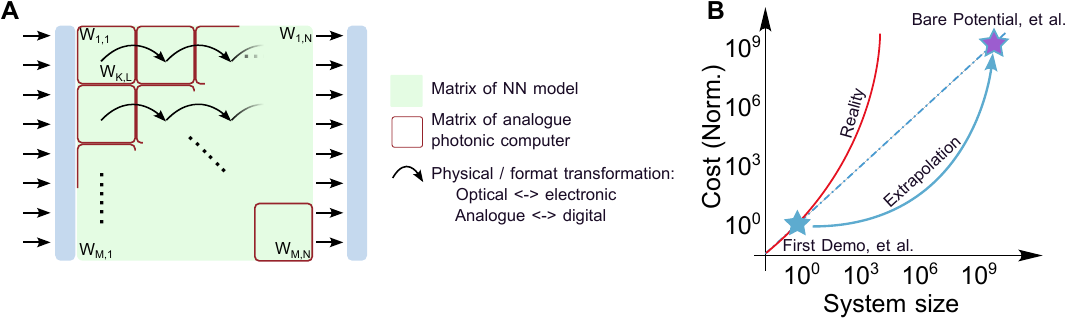}  
	\caption{\textbf{| Why large-range extrapolation of ONN performance is problematic.}
	 \textbf{A} Current matrix vector multiplication (MVM) engines are far too small to implement all weights of an artificial neural network’s (ANNs) layer.
	 ANN matrices are hence tiled into quadratically many small analog MVM, and the costly physical domain and format transformations (black arrows) now scale with $O(N^{2})$.
	 \textbf{B} The true cost and the scaling of these overheads is challenging to estimate, and hence performance extrapolations derived from highly simplified systems tend to strongly underestimate them.
	 Full system validation at scale is required to establish a first true reference for ONN computation.}
	\label{fig:Excitable}
\end{figure*}

\section{Feasibility of common claim and key challenges}

The potential benefits of optical ANN hardware are therefore clear, as is the need for continued innovation and long-term maturation of concepts and systems.
 Achieving this will require continuous support that is justified by relevant goals in the context of potential future applications. 
 In order to match future expectations, the following points should be considered by researchers and policy makers alike.

MVM using passive optics is often presented as a way to reduce the electronic $O(N^2)$ MVM scaling to an $O(N)$ cost.
 While photonics is less energy-efficient than electronics for the $O(N)$ operations interfacing optical, electrical, digital and analogue domains, c.f. black arrows in Fig.~\ref{fig:Excitable}\textbf{A}, the favorable $O(N)$ versus $O(N^2)$ scaling is typically argued to outweigh this drawback.
 However, this neglects a crucial limitation: the size of current integrated photonic MVMs remains far below the MVM- dimensionality used in state-of-the-art ANNs.
 However, tiling an ANN $(N\times N)$ MVM into a smaller $(L\times L)$ photonic MVM requires $(N/L)^2$ tiles, see Fig.~\ref{fig:Excitable}\textbf{A}, and hence the system-level cost for tiling MVMs becomes $O((N/L)^2)$.
 Even worse, each tiling involves interfacing operations at which photonics usually fairs particularly bad, and hence these now scale with $O((N/L)^2)$.
 This largely overlooked fact threatens to nullify the apparent $O(N)$ vs. $O(N^2)$ energy advantage of optical computing - unless integrated photonic platforms can achieve MVM dimension $L$ approaching $N$.

Energy efficiency claims are often made at the level of individual components or for highly simplified system configurations.
 Such claims must be compared fairly with electronic counterparts of equivalent complexity.
 Extrapolating low system-complexity proof-of-concept results to full system performance remains speculative.
 Frequently, such extrapolations overlook the necessary auxiliary hardware; even when included their contribution remains speculative until optical accelerators or ONNs are demonstrated under application conditions within a fully integrated, operational architecture.
 Figure~\ref{fig:Excitable}\textbf{B} highlights this issue in the present context.
 In a recent initiative, large consortia took first steps toward such system-level validation \cite{ahmed2025universal}.
 However, the reported energy efficiencies remain several orders of magnitude below those claimed for system-level optical accelerators derived from extrapolations.

Speed of ONNs or optically accelerated ANNs is often described as unattainable for electronics, with photonics ONNs offering an astonishing potential for ultra-high- speed operation reaching THz bandwidths \cite{fischer2023neuromorphic}.
 While true in principle, such speeds can only be realized if input data is delivered at these rates and in real time; otherwise, the application-relevant inference rate remains limited by the data-input bandwidth, which for ONNs in many and for ANN-acceleration in effectively all cases is provided electronically through opto-electronic modulators.
 Claims of ultra-high bandwidth must therefore be tempered by what is practically achievable for input speeds for input data and input to MVMs from tiling or from previous layers.
 Assertions that single-channel $>100$~GHz electro-optical modulators can be scaled to be massively deployed in large systems often overlook (i) a concrete photonic integration strategy towards such large numbers, (ii) the additional energy costs of the full system implementation, and (iii) the absence of electronic bus architectures capable of driving these modulators at scale with the required throughput.

‘Connectionist’ claims derived from individual, highly specialized photonic com- ponents are often extrapolated to the performance of an entire optical ANN accelerator or ONN.
 In practice, however, properties that make individual components highly performant often do not meet the requirements when applied at scale in large ONNs.
 A prominent example is photonic nonlinearity, which in cutting-edge resonators can approach single-photon energy levels.
 This performance is achievable only with extremely high-quality factors resulting in extraordinary spectral selectivity, making coupling among many such resonators practically impossible due to fabrication non-idealities, temperature gradients, and drifts.
 Even if such integration into networks was feasible, the additional overhead originating from stabilization and control of these devices would outweigh any benefits gained from the exceptional performance of individual components.
 On a more fundamental level, if a computational task requires precision exceeding $10^{-4}$, digital encoding fundamentally outperforms analogue encoding due to thermodynamic limits \cite{boahen2017neuromorph}, negating the primary motivation for the usual analogue photonic ONNs.

Cost of training is almost always overlooked. ONNs, by design, are often not compatible with standard ANN training based on error back-propagation.
 Black-box, model-free, random projection and in-situ learning approaches aiming at a hardware equivalent to error back propagation offer promising alternatives \cite{skalli2025model}.
 Yet the energy cost of such training and how these scale with ONN size remains unexplored.
 Photonic ANN accelerators, meanwhile, operate in a rapidly evolving technological environment.
 In practice, training and inference may incur costs of similar magnitude, and optical ANN accelerators usually require additional effort during training to account for lower resolution and device drift.
 Thus, statements like “train once, exploit for a long time” tend to oversimplify the actual cost dynamics.

\section{Towards true technological relevance}

In order to leverage the synergy between optics and ANN concepts, we propose several strategies spanning multiple domains, each addressing the roadblocks outlined in the previous section.

For optical ANN acceleration, a crucial objective towards mitigating the tiling-induced $O(N^2)$ scaling is to prioritize the development and demonstration of substantially larger, potentially unconventional optical MVM engines instead of optimizing the performance or functionality of individual elements or systems already demonstrated to scale poorly.
 The significant recent progress in integrated optical MVM engines still resulted in severe dimensionality limitations, and one can assume that these are likely to persist without exploration of alternative pathways.
 Achieving breakthroughs in demonstrating optical MVM’s value in practical application scenarios will depend on engineering high-performance digital-electronic to optical interfaces that do not abolish the advantages gained from optics.
 To overcome current inference- rate limitations for electrical data injection, new methods for fast and high-dimensional electro-optical modulation, such as those reported in \cite{panuski2022full}, merit intensified research efforts.

For specialized ONN architectures, increased emphasis should be placed on architectures that implement the full system comprising data injection, nonlinear transformations, and readout mechanisms \cite{skalli2025model} in a synergistic and programmable manner.
 Making such systems accessible to a broad range of applications will also require exploring training strategies tailored to physical ONNs \cite{momeni2025training}.

Free-space optical systems currently represent the state of the art in addressing several of these architectural bottlenecks.
 However, persistent concerns over fabrication costs, long-term stability, and compactness make their relevance for economically viable systems uncertain.
 2D photonic integration may require extending classical approaches to overcome the up-scaling challenges inherent in mapping dense networks onto a 2D topology.
 Techniques such as wavelength-division multiplexing can alleviate congestion only on a shared bus, yet promising strategies could harness rf-multiplexing \cite{dong2023higher}.
 Nevertheless, each hardware-implemented ANN parameter demands a physical instantiation, and thus $O(N^2)$ area scaling persists in classical 2D integration, although recent work on holographic waveguides \cite{dong2023higher} demonstrated the potential to fundamentally improve 2D-integration scalability to $O(N^{1.5})$ within a finite range of $N$ determined by material properties.
 However, in general, 3D integration so far remains the only strategy capable of enabling scalable integration of densely connected networks \cite{boahen2022dendrocentric}.
 Noteworthy, optical 3D integration \cite{moughames2020three} overcomes a fundamental bottleneck of electronic circuits: heat generation in 3D electronics scales with the circuit volume, creating a runaway problem as heat dissipation depends on a circuit’s sur-face area.
 In photonic 3D circuits, scattering and not absorption dominates losses, and active components as the main source of heat can be placed at the surface for efficient heat dissipation, hence heat generated and its dissipation scale equally with system size.
 However, 3D photonic integration strategies remain in their infancy.
 They have not been demonstrated or tested at scale, still suffer from excessive losses, and are not reconfigurable post-fabrication on the required scale.

\section{Conclusion}

As the societal significance and widespread adoption of ANNs continue to grow, concerns about their scalability and energy consumption become central, underscoring the need for solutions that utilize the strengths of optics.
 Indeed, both Nvidia and TSMC are now officially exploring photonic integration, yet so far, the goal remains limited to bringing optical communication in data centers closer to electronic digital compute.
 In this context, ONNs find themselves attracting a level of interest hitherto unmatched, and the implementation of large-scale ONN computing promises to be transformative.
 Current claims are often based on idealized component-level results rather than full system performance, and when realistic architecture requirements are included, much of the projected scaling benefit tend to vanish.
 Present photonic technology cannot yet support the large matrix sizes used in state-of-the-art AI, and breaking computations into smaller optical blocks re-introduces the same $O(N^2)$ overhead that photonics aims to eliminate.
 Likewise, ultra-high-speed components are only useful if data can be fed at comparable rates, which remains an unsolved bottleneck.
 Finally, breakthroughs in nonlinearity or efficiency at the component level rarely translate into scalable and reliable networks, while photonic and analogue-specific training costs are frequently underestimated or even ignored.
 Credible progress demands (i) system-level validation under realistic ANN conditions, (ii) technology roadmaps explicitly targeting large-scale MVM dimensionality, high-bandwidth data injection, co-designed nonlinear and digital interfacing, and (iii) application specific key performance indicators distinguishing general-purpose acceleration from more specific physical ONN computing.
 Some very encouraging recent advances highlight the potential of ONNs for complex and large-scale ANN tasks of practical relevance, including error-tolerant photonic architectures enabling networks over 100 layers deep at high data rates and demonstrating natural language processing and large-scale computer vision \cite{zhou2025hundred}.
 
Despite significant progress and the wide range of approaches aiming to circum-vent current limitations to reach practical optical computing, delivering photonic ANN computing will likely neither be a short-term race nor a repetition of past optical-computing cycles.
 It requires cooperation between researchers, funders, industry, as well as strengthening the communication pipeline between the photonic ONN com-munity and potential application domains and engineers. Developing such potentially long-term programs towards true progress and relevance requires trust that such an endeavor is, both, possible and worthwhile.
 If conditions are created, then ONNs may ultimately progress from elegant proof-of-concept demonstrations to indispensable technologies.
 The path forward remains narrow, but the scientific, economic, and societal potential remain sufficiently compelling to justify the journey.

\section{Acknowledgments}
	
We acknowledge support from the European Union’s Horizon 2020 research and innovation programme with European Research Council (ERC) (Consolidator Grant, grant agreement no. 101044777 (INSPIRE)), with the Marie Skłodowska-Curie grant agreement No 860830 (POST DIGITAL) and No 101169118 (POSTDIGITAL+), by the French Agence Nationale de la Recherche (ANR), by the French Investissements d’Avenir program, project ISITE-BFC (contract ANR-15-IDEX-03) and EUR EIPHI program (Contract No. ANR-17-EURE-0002) as well as support by the Region Bourgogne Franche Comte.

\section{References}

\bibliography{bibliography}
	
\end{document}